\documentclass[review]{elsarticle}
\usepackage{lineno,hyperref}
\modulolinenumbers[5]
\usepackage{geometry}
\journal{Journal of \LaTeX\ Templates}
\usepackage{verbatim}
\usepackage{amsfonts}
\usepackage{graphics}
\usepackage{subfigure}
\usepackage{float}
\usepackage{tikz}
\usepackage{booktabs}  
\usepackage{threeparttable}  
\usepackage{makecell}
\usepackage{multirow}
\usepackage{multicol}
\usepackage{appendix}
\usepackage{color}
\usepackage{amsmath}
\usepackage{verbatim}
\usepackage{algorithm}
\usepackage{algorithmic}

\biboptions{authoryear}

\begin{document}

\begin{frontmatter}
\title{Dynamic mapping from static labels: remote sensing dynamic sample generation with temporal-spectral embedding}


\author[HKU,PCL]{Shuai Yuan}
\author[HKU,PCL]{Shuang Chen}
\author[HGS,PCL]{Tianwu Lin}
\author[PCL]{Jincheng Yuan}
\author[PCL]{Geng Tian}
\author[PCL]{Yang Xu}
\author[PCL]{Jie Wang}
\author[HKU]{Peng Gong\fnref{myfootnote1}}



\address[HKU]{Department of Geography, The University of Hong Kong, Hong Kong, China}
\address[PCL]{Pengcheng Laboratory, Shenzhen, China.}
\address[HGS]{Department of Electronics and Information Engineering, Harbin Institute of Technology (Shenzhen), Shenzhen, China}

\fntext[myfootnote1]{Corresponding author: penggong@hku.hk}

\begin{abstract}\label{abstract}
Accurate remote sensing geographic mapping requires timely and representative samples. However, rapid land surface changes often render static samples obsolete within months, making manual sample updates labor-intensive and unsustainable. To address this challenge, we propose TasGen, a two-stage Temporal spectral-aware Automatic Sample Generation method for generating dynamic training samples from single-date static labels without human intervention. Land surface dynamics often manifest as anomalies in temporal-spectral sequences. 
To effectively capture these dynamics, TasGen first disentangles temporal and spectral features to isolate their individual contributions, and then couples them to model their synergistic interactions. In the first stage, we introduce a hierarchical temporal-spectral variational autoencoder (HTS-VAE) with a dual-dimension embedding to learn low-dimensional latent patterns of normal samples by first disentangling and then jointly embedding temporal and spectral information. This temporal-spectral embedding enables robust anomaly detection by identifying deviations from learned joint patterns. In the second stage, a classifier trained on stable samples relabels change points across time to generate dynamic samples. To not only detect but also explain surface dynamics, we further propose an anomaly interpretation method based on Gibbs sampling, which attributes changes to specific spectral-temporal dimensions. We evaluate TasGen on two dynamic geographic mapping tasks (i.e., land cover mapping with 5 classes and wetland mapping with 4 classes) across six diverse global locations, each covering an area of around $12,062 \ km^2$ and representing different continents and climate zones. TasGen achieved an average mapping OA and Kappa of 91.64\% and 0.9071, considerably outperforming other state-of-the-art algorithms with a maximum of 7.97\% and 0.0762. Consistent superior accuracy across all regions demonstrates TasGen’s effectiveness in tracking both long-term trends and abrupt land surface changes, while significantly reducing the need for human intervention.


\end{abstract}



\end{frontmatter}

\section{Introduction}\label{intro}
%
The Earth's surface is undergoing unprecedented rapid changes, with urbanization encroaching upon natural habitats at an annual rate of 1.2\% \citep{seto2012global}, and global wetlands decreasing by approximately 0.5\% per year since the 20th century \citep{nicholls2004coastal, yuan2025comprehensive}. Timely and accurate geographic mapping serves as a fundamental basis for monitoring ecosystem health \citep{soubry2021systematic, mariano2018use}, optimizing urban planning \citep{gong2020mapping, yuan2024fusu, yuan2020long}, protecting biodiversity \citep{wiens2009selecting}, and understanding climate change \citep{yang2013role}. Remote sensing has become the primary way of acquiring dynamic surface information with its capability for periodic observations \citep{gong2012remote, pekel2016high}. For instance, the Landsat satellite series has provided over 40 years of continuous observations, offering essential data for global land change analysis \citep{wulder2019current}.

However, high-precision remote sensing mapping relies heavily on high-quality training samples that are both representative and temporally relevant \citep{stanimirova2023global}. Main approaches depend on manually labeled samples to construct training samples, yet the nonlinear nature of land surface changes (e.g., the multi-phase expansion of urban areas, the stochastic nature of wetland variations) renders these samples obsolete within months \citep{winkler2021global, lin2018losses}. For example, if the change rate of built-up areas in a city from T1 to T2 reaches over 20\%, and a classification model trained on samples labeled in T1 cannot guarantee stable classification when applied to T2 data \citep{gong2019stable, gong2024stable}. Even though at the global scale, the annual change rate is less than 4\%, yet both regional wetland ecosystems and urban areas can see change rates over 20\% in one year \citep{davidson2014much, gao2020mapping}. As a result, this "labeling lag" is a critical bottleneck limiting the effectiveness of dynamic remote sensing monitoring. 

Now there are three kinds of methods addressing this challenge. The first is manually annotating multi-temporal samples. It requires visual interpretation and manual annotations at each time point, which is constrained in small study areas and short time series easily. For example, \citet{li2017first} established the first all-season sample sets (FAST) for global land cover mapping, primarily based on manual interpretation of Landsat-8 imagery regarding high-resolution Google Earth images. \citet{essd-13-3907-2021} collected training samples by integrating land cover samples derived from China’s Land-Use/Cover Datasets (CLUDs) with visually interpreted samples based on satellite time-series data, Google Earth, and Google Maps imagery. The second is using stable samples across time series to train phase-general classifiers. In other words, these methods only need manual annotations once to conduct the time-series mapping. For example, \citet{zhang2016annual} developed a method for extracting training samples from spectrally stable areas, and used Google Earth imagery as reference data for land cover identification. \citet{huang2020migration} developed a novel methodology to migrate global training samples by leveraging spectral similarity and spectral distance metrics, enabling the reuse of stable training samples for dynamic global land cover mapping across multiple periods. Even though stable samples only need one-time annotation, the scarcity of stable samples and the temporal identification of stable samples limit the scaling ability. The third is the change detection method modeling the time-series patterns. \citet{zhu2014continuous} proposed the continuous change detection and classification (CCDC) algorithm, detecting changes through unsupervised time-series modeling and training classifiers relying on static reference data from stable periods. \citet{kennedy2010detecting} proposed LandTrendr, considering dynamic segmentation analysis of Landsat time series and training classifiers relying on trajectory fitting and seasonal-trend decomposition. \citet{he2024time} employed temporal semantic segmentation to extract land cover change information from Sentinel-2 time series data. These methods directly identify change points and their types at the mapping level, modeling temporal dynamics in an end-to-end manner. While powerful, \citet{zhu2014continuous} and \citet{kennedy2010detecting} struggle to adapt to serious Earth's surface patterns because of the model initialization. Besides, they only considered the temporal dimension but neglected the synergistic effects of temporal-spectral patterns when land surface changes. \citet{he2024time} has limited scalability because of the need for temporal-dense annotations. Additionally, such frameworks can benefit from the integration of intermediate constraints at the sample level, with better interpretation ability and enhanced control over the learning process. 

\begin{figure}[h]
    \centering
    \includegraphics[width=0.85\linewidth]{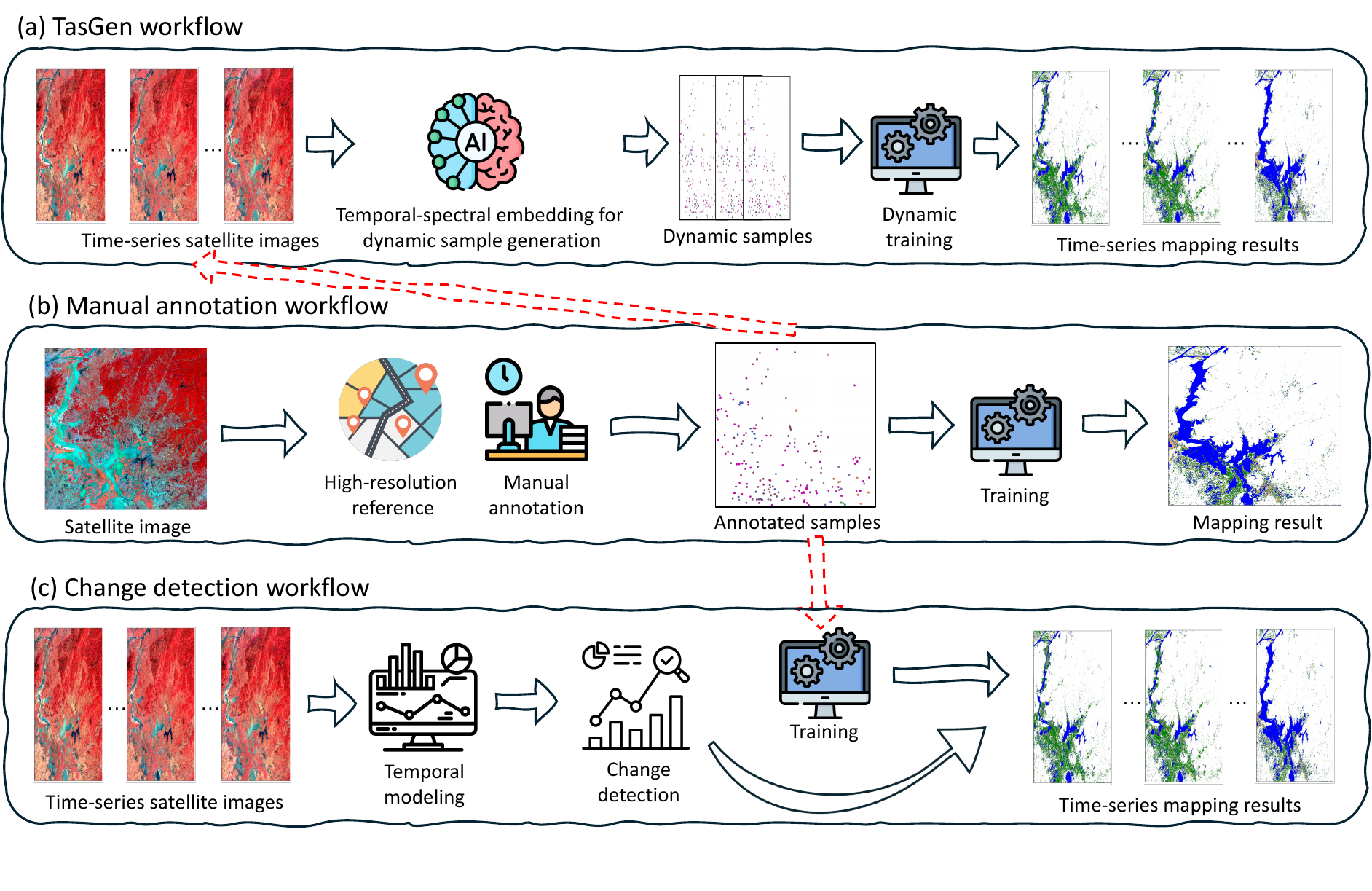}
    \vspace{-1em}
    \caption{The workflow of each dynamic mapping paradigm. (a) is the workflow of our proposed method; (b) is the workflow of the manual annotation; (c) is the workflow of the change detection methods. We introduce dynamic samples as an interpretable and controllable intermediate variable for final classification results.}
    \label{fig:workflow}
\end{figure}

Due to the complexity of both spectral and temporal dynamics inherent in time-series data, obsolete samples cannot effectively represent surface characteristics at later time points, leading to significant degradation in mapping accuracy over time. High-quality and abundant dynamic and up-to-date samples are the key to accurate time-series dynamic mapping \citep{stanimirova2023global}. Based on the observations above, dynamic samples are needed, but manual annotations are labor-consuming, and neglect the inherent temporal dynamics, impractical for continuous monitoring of large-scale geographical areas experiencing rapid surface alterations (Fig. \ref{fig:workflow} (b)). Existing change detection methods often rely on strong assumptions about temporal patterns or temporal-dense annotations, and neglect the synergistic effects of temporal-spectral patterns. This limitation is particularly problematic for wetlands, which exhibit recurrent phenological cycles, and subtle intra-annual transitions that are only distinguishable through the joint modeling of spectral signatures and their temporal evolution \citep{yuan2025comprehensive} (Fig. \ref{fig:workflow} (c)). Besides, while end-to-end frameworks offer efficiency, introducing intermediate sample-level constraints allows for greater flexibility and interpretability, improving robustness in diverse and evolving environments. Therefore, we argue that a more robust approach involves initially performing change detection on individual training samples to get dynamic samples across time series. By introducing dynamic samples as an interpretable and controllable intermediate variable, final classification results are ensured to be grounded in well-verified surface transition evidence. These dynamic samples can then be employed to train classifiers that yield improved dynamic mapping results (Fig. \ref{fig:workflow} (a)). This two-stage approach provides greater control over intermediate variables, significantly enhancing the accuracy of dynamic change detection outcomes.

\begin{figure}[h]
    \centering
    \includegraphics[width=0.8\linewidth]{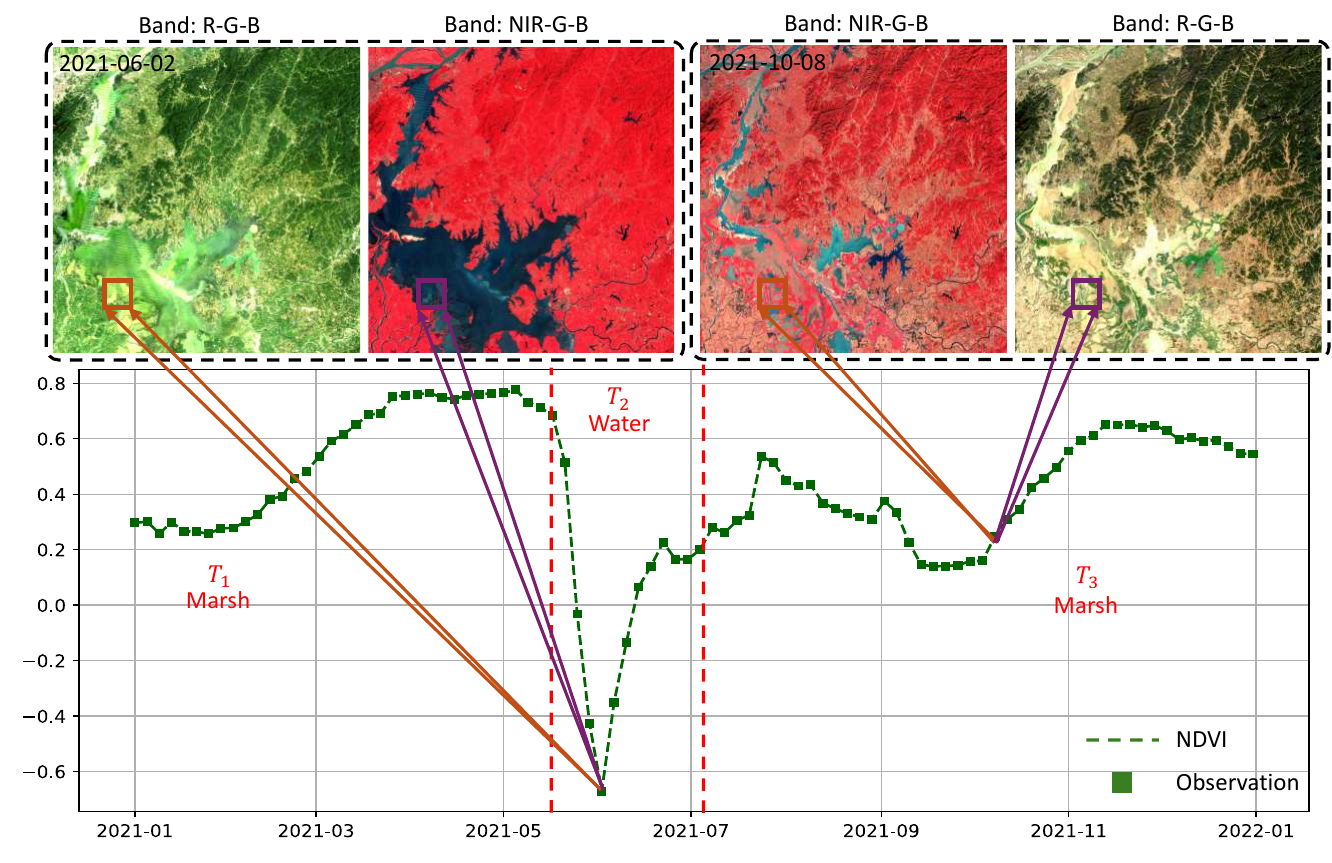}
    \vspace{-1em}
    \caption{The land surface change is a form of spatiotemporal anomaly. This set of figures and the NDVI variation show how the land surface type in the rectangle transformed. In May-June 2021, the inland marsh transformed to surface water, representing a spatiotemporal anomaly in NDVI.}
    \label{fig:anomaly_1}
\end{figure}

Fundamentally, land surface change (e.g., deforestation, urban expansion, wetland degradation) can be understood as a form of spatiotemporal anomaly. As Fig. \ref{fig:anomaly_1} shows, these changes are typically characterized by disruptions in temporal trends and simultaneous deviations across multiple spectral dimensions. For instance, the breakdown of expected correlations between bands (e.g., NIR and SWIR) or the disappearance of seasonal vegetation cycles in indices like NDVI often signal such changes. Recognizing these patterns as spatiotemporal anomalies provides a conceptual foundation for detecting land surface dynamics in time-series remote sensing data. Therefore, a key step towards robust dynamic geographic mapping is to generate reliable dynamic training samples from limited static labels by identifying such anomalies. However, generating dynamic samples from static supervision faces two main challenges.
First, remote sensing time-series dynamics are both multivariate and unified, which means they involve distinct temporal and spectral components that reflect different aspects of surface change, yet jointly characterize a coherent land surface state. Purely decoupled learning of temporal and spectral embeddings may fail to capture their integrated semantic meaning, while purely coupled modeling risks entangling fundamentally different variation sources. This tension makes it challenging to achieve consistent and effective temporal-spectral representation learning. Second, the lack of interpretable attribution in anomaly detection limits the model’s ability to precisely localize the source of changes. Without analyzing the contributions of different spectral bands and time points, anomalies may be detected without a clear causal explanation, reducing model robustness and interpretability, especially in complex or heterogeneous environments.

To generate reliable dynamic training samples under such constraints, we propose a two-stage Temporal-spectral-aware Automatic Sample Generation method (TasGen). At the first stage, TasGen identifies the normal and anomalous spectral and temporal patterns of samples in a period by encoding time-series data into low-dimensional embeddings in an unsupervised manner. We first propose a dual-dimension embedding to decouple and compress the dynamic time series along both temporal and spectral dimensions to obtain the low-dimensional temporal and spectral embeddings. We then develop a hierarchical temporal-spectral Variational AutoEncoder (HTS-VAE) with the two stochastic latent variables to jointly learn the low-dimensional temporal and spectral embeddings. 
At the second stage, TasGen detects and relabels abnormal temporal samples via a trainable classifier, dynamically generating the training samples across different periods without extra manual annotation. Moreover, we develop an anomaly interpretation method based on Gibbs sampling that iteratively imputes suspected anomalous bands and time points. By approximating the normal reconstruction pattern, this approach allows precise attribution of anomalies to specific spectral-temporal components, enhancing the model's robustness and interpretability in complex surface change scenarios. 

The contributions of this paper are summarized as follows:
\begin{itemize}
    \item Our proposed TasGen is the first method focusing on automatic dynamic sample generation for time-series dynamic remote sensing mapping. By first decoupling temporal and spectral dependencies and then jointly modeling them via a dual-dimension embedding and a hierarchical temporal-spectral Variational AutoEncoder (HTS-VAE), TasGen aligns and models patterns across spectral and temporal dimensions to enable accurate anomaly detection.
    \item To achieve precise interpretable attribution, TasGen designs a Gibbs-based anomaly interpretation method, which effectively filters transient anomalies and attributes changes to specific spectral-temporal dimensions.
    \item Two comprehensive evaluation experiments, including dynamic land cover mapping and wetland mapping across six different regions, demonstrate the effectiveness and robustness of our proposed TasGen, confirming TasGen's potential to enhance automated geographical mapping practices and reduce the associated labor demands in remote sensing dynamic monitoring.
\end{itemize}

\section{TasGen}

\subsection{Problem definition}
We consider a single time-series dynamic sample $\mathcal{S} \in R^{C, N}$ as an example, where $C$ and $N$ are the numbers of bands and observations of the time-series remote sensing data, respectively. We already know the type $\mathcal{y}_t$ of this sample at time $t$, and our final task is to automatically determine all the types of this sample across the observations $N$ based on this existing annotation and the time-series dynamic patterns. We see the patterns similar to the time-series data centered around the known $t$ are normal. If there is a type change, temporal-spectral changes must occur, leading to data instances significantly deviating from the normal observations. Based on this understanding, we then can split the task into three sub-tasks: First, we need to identify whether an observation $\mathbf{x}_t$ is an anomaly or not via the anomaly detector $\mathcal{A}$; Second, we should analyze the contributions of different spectral bands at various time points to the anomaly detection result, enabling precise localization of change sources and enhancing the overall robustness and interpretability of the method; Third, for the anomaly, we determine the type of anomaly via a classifier $\mathcal{C}$, and for the normals, we keep the type $y_t$ (shown in Equ. \ref{equ:1}); 

\begin{equation}  
{y}_t = \left\{  
  \begin{array}{lr}  
    {y}_t \ \ , &\ \ \mathcal{A}(\mathbf{x}_t) = 0  \\[8pt]
    \mathcal{C}(\mathbf{x}_t) \ \ , &  \mathcal{A}(\mathbf{x}_t) = 1 \\[8pt]    
  \end{array}  
\right.  
\label{equ:1}
\end{equation}

\subsection{Overview of TasGen}

\begin{figure}[h]
    \centering
    \includegraphics[width=0.95\linewidth]{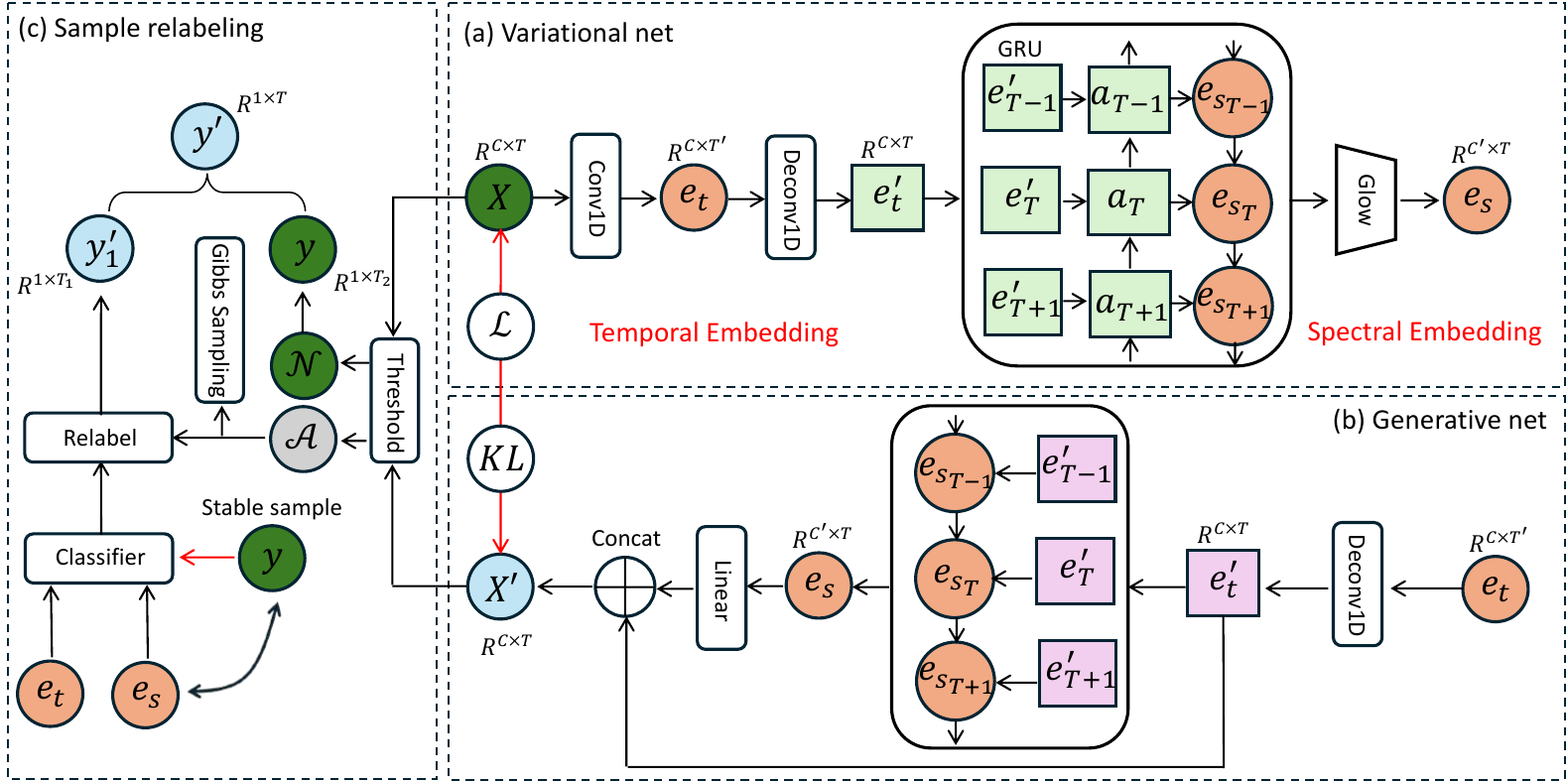}
    \vspace{-1em}
    \caption[The overall architecture of TasGen.]{
The overall architecture of TasGen. (a) is the variational net, and (b) is the generative net, and (c) is the sample relabeling process. \protect\tikz[baseline=(X.base)]{
  \node[draw=black, fill=green!48!black, circle, inner sep=2pt] (X) {\scriptsize $X$};
} is the input sample with a shape of $C$ bands and $T$ time points. \tikz[baseline=(E.base)]{
  \node[draw=black, fill=orange!60, circle, inner sep=2pt] (E) {\normalsize $\mathit{e_t}$};
} is the temporal embedding, and \tikz[baseline=(E.base)]{
  \node[draw=black, fill=orange!60, circle, inner sep=2pt] (E) {\normalsize $\mathit{e_s}$};
} is the spectral embedding. 
\protect\tikz[baseline=(X.base)]{
  \node[draw=black, fill=cyan!20, circle, inner sep=1.8pt] (X) {\scriptsize $X'$};
} represents the reconstructed time-series patterns of the dynamic sample. \tikz[baseline=(KL.base)]{
  \node[draw=black, fill=white, circle, inner sep=1pt] (KL) {\normalsize $\mathit{KL}$};
} is the KL divergence. \tikz[baseline=(L.base)]{
  \node[draw=black, fill=white, circle, inner sep=2pt] (L) {\normalsize $\mathcal{L}$};
} is the ELBO loss. \tikz[baseline=(Y.base)]{
  \node[draw=black, fill=green!48!black, circle, inner sep=2.5pt] (Y) {\normalsize $\mathit{y}$};
} is the label of stable samples. \tikz[baseline=(N.base)]{
  \node[draw=black, fill=green!48!black, circle, inner sep=2pt] (N) {\normalsize $\mathcal{N}$};
} and \tikz[baseline=(A.base)]{
  \node[draw=black, fill=gray!48, circle, inner sep=2pt] (A) {\normalsize $\mathcal{A}$}
} mean the normal patterns and the anomalies. 
 \tikz[baseline=(Y.base)]{
  \node[draw=black, fill=cyan!20, circle, inner sep=2pt] (Y) {\normalsize $\mathit{y'}$};
} is generated labels for the dynamic sample.  }

    \label{fig:net}
\end{figure}

TasGen is a two-stage framework designed to automatically generate dynamic training samples for remote sensing dynamic geographic mapping. The overall framework is illustrated in Fig. \ref{fig:net}. Given a static labeled sample at a specific time point, TasGen aims to reconstruct its temporal evolutions and generate the dynamic states across the full observation window, capturing underlying temporal-spectral patterns without additional manual annotation.

To this end, TasGen first decouples the temporal and spectral dependencies through a dual-dimension embedding, and then learns disentangled spectral and temporal representations jointly through a hierarchical temporal-spectral autoencoder (HTS-VAE). Anomalies are detected based on the reconstructed temporal-spectral patterns. In the second stage, dynamic sample relabeling is conducted using a classifier trained by the learned embeddings, enabling temporal label propagation from a single static annotation. Additionally, we introduce a Gibbs-based anomaly attribution mechanism to interpret the contributions of spectral-temporal components and enhance model transparency.

\subsection{Dual-dimension embedding}
We aim to model the temporal-spectral patterns of remote sensing time-series dynamic samples to detect anomalies, contributing to generating dynamic remote sensing samples. 
To explicitly model the dependencies, we first propose a dual-dimension embedding architecture, which disentangles temporal and spectral representations by compressing the time-series input along both the temporal and spectral dimensions. 

\begin{figure}[h]
    \centering
    \includegraphics[width=0.5\linewidth]{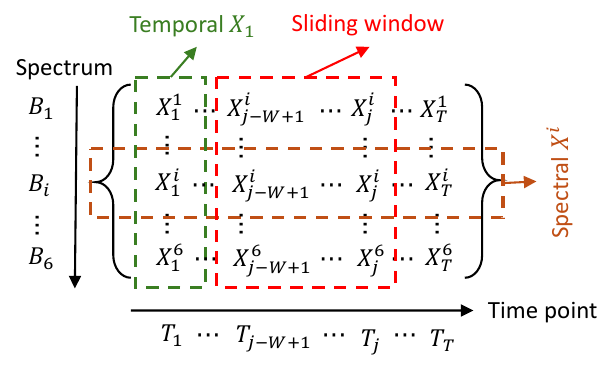}
    \vspace{-1em}
    \caption{The sample data structure.}
    \label{fig:window}
\end{figure}

As shown in Fig. \ref{fig:window}, we normally use a sliding window over the time-series data to calculate and model. Given an input window $\textbf{x}_{1:T} \in \mathbb{R}^{C \times T}$ with $C$ spectral bands and $T$ time points, we first apply stacked Conv1D layers $f$ along the temporal dimension:
\begin{equation}
\textbf{e}_t = f(\textbf{x}_{1:W}) \in \mathbb{R}^{C \times T'}, \quad T' = T / 2
\end{equation}

This produces a compressed temporal embedding $\textbf{e}_t$ that captures temporal-spectral dynamics such as phenological cycles or disturbance trends. The embedding is then decoded via Deconv1D $D$ layers to reconstruct a smoothed version of the input:
\begin{equation}
\textbf{e}'_{t} = D(\textbf{e}_t) \in \mathbb{R}^{C \times T}
\label{formula2}
\end{equation}

Static annotations often reflect transient surface states, risking model overfitting to anomalous spectral-temporal patterns. To address this, we adopt a sample curation strategy: instead of directly learning from raw inputs, we first reconstruct a stable time series $\textbf{e}'_t$ using the temporal embedding $\textbf{e}_t$, then derive spectral embeddings from this denoised signal. This approach preserves typical surface dynamics while filtering short-term noise (e.g., clouds), thereby reducing annotation bias. We further pretrain the reconstruction path using a temporal VAE to ensure early-stage stability and improve generalization.
We then extract the spectral embedding $e_s$ by applying a spectral-wise recurrent encoder (bidirectional GRU $G$ \citep{chung2014empirical}) along the spectral dimension:
\begin{equation}
\textbf{e}_s = G(\textbf{e}'_{t}) \in \mathbb{R}^{C' \times T}, \quad C' = C / 2
\end{equation}
This embedding captures coordinated spectral responses indicative of surface properties (e.g., wetland spectral signatures across NIR, SWIR, and Red).

Our dual-dimension embedding decouples temporal and spectral encoding. This separation preserves the alignment between latent representations and the input sequence, ensuring that inter-spectral dependencies are learned at the correct temporal granularity.

\subsection{Hierarchical temporal-spectral variational autoencoder}

While our dual-dimension embedding structure effectively disentangles temporal and spectral representations, these two embeddings are learned independently. This separation may result in a lack of contextual interaction between temporal and spectral embeddings. To address this, we introduce a hierarchical variational autoencoder (HTS-VAE) (Fig. \ref{fig:net} (a)(b)) that jointly models these two embeddings in a coherent latent space. Specifically, we treat the temporal embedding as a contextual prior for spectral embedding learning, thereby enabling temporal-spectral simultaneous learning while preserving the structural benefits of disentangled encoding.

HTS-VAE utilizes two hierarchical stochastic latent variables $\textbf{e}_t$ and $\textbf{e}_s$ to represent the sample with the time-series sequence $x \in \mathbb{R}^{C \times T}$ with $C$ spectral bands and $T$ time points. The joint representation model can be abstractly defined as:
\begin{equation}
p_\theta(\textbf{x}, \textbf{e}_s, \textbf{e}_t) = p_\theta(\textbf{x} \mid \textbf{e}_s, \textbf{e}_t) \cdot p_\theta(\textbf{e}_s \mid \textbf{e}_t) \cdot p_\theta(\textbf{e}_t).
\end{equation}

The encoder jointly learns the temporal embedding $\textbf{e}_t$ and the spectral embedding $\textbf{e}_s$, ensuring that $\textbf{e}_s$ captures band-wise relationships conditioned on temporal context $\textbf{e}_t$. The prior distribution $p_\theta(\textbf{e}_t)$ is assumed to be Gaussian, forcing the latent variables to learn stable temporal patterns. It provides a probabilistic framework for learning temporal patterns, constraining the range and characteristics of $\textbf{e}_t$ to guide the model towards stable time-aware embeddings. The conditional distribution $p_\theta(\textbf{e}_s \mid \textbf{e}_t)$ ensures that spectral information is aligned with temporal context. The learning of spectral embedding is thus contextualized by the learned temporal patterns. This avoids the challenge of fusing independently learned $\textbf{e}_t$ and $\textbf{e}_s$. The likelihood $p_\theta(x \mid \textbf{e}_s, \textbf{e}_t)$ is the decoder, which describes how to reconstruct the high-dimensional input $\textbf{x}$ from the two latent embeddings. During training, the model maximizes this likelihood to ensure that the reconstructed data closely resembles the input. By optimizing this, the model learns to integrate temporal and spectral dependencies into the reconstruction process, effectively capturing the normal pattern of samples.

The variational posterior can be factorized as:
\begin{equation}
q_\phi(\textbf{e}_s, \textbf{e}_t \mid \textbf{x}) = q_\phi(\textbf{e}_s \mid \textbf{e}_t) \cdot q_\phi(\textbf{e}_s \mid \textbf{x}).
\end{equation}
Specifically, as Fig. \ref{fig:net} (a) shows, we use an SRNN-list structure \citep{fraccaro2016sequential} to deduce a more powerful spectral embedding that is aware of temporal context. It means, at each time point $t$, $q_\phi(\textbf{e}_{s_t} \mid \textbf{e}_{s_{t-1}}, \textbf{a}_t = h_{\phi_{\textbf{a}}}(\textbf{a}_{t+1}, \textbf{e}'_{t_t})),$ where $\textbf{a}_t$ is the backward GRU state derived from the backward GRU Function $h_{\phi_{\textbf{a}}}(\cdot)$. $\textbf{e}'_{t_t}$ is the curated temporal embedding, used to calculate the next dependency $q_\phi(\textbf{e}_{s_{t}} \mid \textbf{e}'_{t_{t+1:T}})$ in the input time series. 

The standard VAE objective is to maximize the evidence lower bound (ELBO):
\begin{equation}
\mathcal{L}(\textbf{x}, \theta, \phi) = \mathbb{E}_{q_{\phi}(\textbf{e} \mid \textbf{x})}[\log p_{\theta}(\textbf{x} \mid \textbf{e})] - D_{\text{KL}}(q_{\phi}(\textbf{e} \mid \textbf{x}) \parallel p_{\theta}(\textbf{e})),
\end{equation}
where $D_{\text{KL}}$ is the Kullback-Leibler divergence and the expectation is estimated using the stochastic gradient variational Bayes (SGVB) estimator. 

In our hierarchical structure, we introduce auxiliary deterministic variables $\textbf{d}$ and $\textbf{z}$ to support the dual-dimension encoding and reconstruction. The training objective becomes:
\begin{equation}
\begin{aligned}
\mathcal{L}(\textbf{x}, \theta, \phi) = & \mathbb{E}_{q_{\phi}(\textbf{e}_s, \textbf{e}_t, \textbf{d} \mid \textbf{x})} \left[ \log p_{\theta}(\textbf{x} \mid \textbf{e}_s, \textbf{e}_t, \textbf{z}) \right] \\
& - D_{\text{KL}}\left( q_{\phi}(\textbf{e}_s, \textbf{e}_t, \textbf{z} \mid \textbf{x}) \parallel p_{\theta}(\textbf{e}_s, \textbf{e}_t, \textbf{z}) \right).
\end{aligned}
\label{eq:elbo_interfusion}
\end{equation}
This can be further expanded as:
\begin{equation}
\begin{aligned}
\mathcal{L}(\textbf{x}, \theta, \phi) = \mathbb{E}_{q_{\phi}} \Big[ & \log p_{\theta}(\textbf{x} \mid \textbf{e}_s, \textbf{e}_t, \textbf{z}) + \log p_{\theta}(\textbf{e}_s, \textbf{e} \mid \textbf{e}_t) \\
& + \log p_{\theta}(\textbf{e}_t) - \log q_{\phi}(\textbf{e}_s, \textbf{d} \mid \textbf{e}_t, \textbf{x}) - \log q_{\phi}(\textbf{e}_t \mid \textbf{x}) \Big],
\end{aligned}
\end{equation}
where $\textbf{d}_{1:T}$ is the deterministic reconstructed input derived from the Deconv1D layer after the temporal embedding $\textbf{e}_t$, and thus $\textbf{d}_{1:T} \sim q_{\phi}(\textbf{d}_{1:T} \mid \textbf{e}_t, \textbf{x}) = q_{\phi}(\textbf{d}_{1:T} \mid \textbf{e}_t) = \delta(\textbf{d}_{1:T} - D(\textbf{e}_t))$, following a delta distribution. Similarly, $p_{\theta}(\textbf{z}_{1:T} \mid \textbf{e}_s) = \delta(\textbf{z}_{1:T} - D(\textbf{e}_s))$. By sharing the parameters of the Deconv1D layer in both encoder and decoder networks, the two delta distributions become equal, i.e., $q(\textbf{d}_{1:T} \mid \textbf{e}_t) = p(\textbf{z}_{1:T} \mid \textbf{e}_s)$, which cancels them out in the ELBO calculation.

To further estimate the ELBO efficiently, we rely on the structural dependencies illustrated in Fig. \ref{fig:net}. The posterior can be reformulated as:
\begin{equation}
\iiint q_{\phi}(\textbf{e}_s, \textbf{e}_t, \textbf{d}_{1:T} \mid \textbf{x}) \ d\textbf{e}_sd\textbf{e}_td\textbf{d}_{1:T} = \iint q_{\phi}(\textbf{e}_s \mid \textbf{d}_{1:T} = D(\textbf{e}_t)) \, q_{\phi}(\textbf{e}_t \mid \textbf{x}) \ d\textbf{e}_sd\textbf{e}_t.
\end{equation}

This formulation allows for a practical two-step Monte Carlo \citep{geweke1989bayesian} estimation to compute the expectations: First, sample $\textbf{e}_t^{(l)} \sim q_{\phi}(\textbf{e}_t \mid \textbf{x})$ for $l = 1, \dots, L$. Then, for each $\textbf{e}_t^{(l)}$, compute $\textbf{d}^{(l)} = D(\textbf{e}_t^{(l)})$, and sample $\textbf{e}_s^{(l)} \sim q_{\phi}(\textbf{e}_s \mid \textbf{d}^{(l)})$. The ELBO expectation terms are then approximated by averaging over these $L$ samples. Since both $q$ and $p$ are parameterized as diagonal Gaussians, most log-probability terms can be computed analytically, which greatly improves training efficiency.

For terms involving $p(\textbf{e}_s, \textbf{z} \mid \textbf{e}_t)$ and $q(\textbf{e}_s, \textbf{d} \mid \textbf{e}_t, \textbf{x})$, we apply the identity $q(\textbf{d}_{1:T} \mid \textbf{e}_t) = p(e_{1:W} \mid z_2)$ and obtain:
\begin{equation}
\begin{aligned}
    \log p_\theta(\textbf{e}_s, \textbf{z} \mid \textbf{e}_t) - \log q_\phi(\textbf{e}_s, \textbf{d} \mid \textbf{e}_t, \textbf{x}) & =  
    \log p_\theta (\textbf{e}_{s_{1: T}} | \textbf{z}_{1: T}) - \log q_\phi (\textbf{e}_{s_{1: T}} | \textbf{d}_{1: T}) \\
    & = \sum_{t=1}^{T} \log p_\theta(\textbf{e}_{s_t} \mid \textbf{e}_{s_{(t-1)}}, \textbf{z}_t) - \log q_\phi(\textbf{e}_{s_t} \mid \textbf{e}_{s_{(t-1)}}, \textbf{a}_t = h_\phi(\textbf{a}_{t+1}, \textbf{d}_t)),
\end{aligned}
\end{equation}
where $\textbf{a}_t$ is the backward GRU state, and $h_\phi(\cdot)$ is the backward GRU Function.
To enhance posterior expressiveness to obtain a more powerful one, we introduce a Glow transformation \citep{kingma2018glow} through an invertible mapping. We have:
\begin{equation}
\textbf{e}_K = f^{-1}_\lambda(\textbf{e}_0), \quad \log q_\phi(\textbf{e}_{s_{t}}^{(K)} \mid \textbf{e}_{s_{t-1}}, \textbf{a}_t) = \log q_\phi(\textbf{e}_{1_t}^{(0)} \mid \textbf{e}_{1_{t-1}}, \textbf{a}_t) + \log \left| \det \frac{\partial f_\lambda(\textbf{e}_{1_t}^{(K)})}{\partial \textbf{e}_{1_t}^{(K)}} \right|.
\end{equation}
Here, $f_\lambda$ is composed of $K$ invertible transformations using Glow-style affine coupling layers. This ensures the posterior $q(\textbf{e}_s \mid \textbf{d})$ has sufficient flexibility to match complex distributions. Substituting the Glow-transformed terms and sequential estimation into the ELBO completes the objective. The model is optimized using the SGVB estimator with reparameterization.

This hierarchical design provides a principled and efficient way to model temporal-spectral patterns, align latent structures with real-world surface dynamics, and robustly support downstream dynamic sample generation and anomaly attribution.

\subsection{Anomaly detection inference}

After learning the joint temporal-spectral representations using HTS-VAE, we perform anomaly detection on dynamic remote sensing samples by evaluating reconstruction-based deviations. 
Given an input window $x \in \mathbb{R}^{C \times T}$, we pass it through the trained HTS-VAE to obtain latent embeddings $e_t$ and $e_s$. These embeddings are decoded to generate a reconstruction $\hat{x}$. We define the anomaly score as the negative log-likelihood (or equivalently, the reconstruction probability):

\begin{equation}
\text{S} = -\mathbb{E}_{q_\phi(\textbf{e}_t, \textbf{e}_s \mid \textbf{x})} \left[ \log p_\theta(\textbf{x} \mid \textbf{e}_t, \textbf{e}_s) \right]
\end{equation}

A higher score indicates lower reconstruction likelihood, and thus, there is a greater chance that the sample contains an anomaly.

\subsection{Anomaly attribution}
To not only detect but also explain land surface dynamics in the dynamic samples, we further need to propose an anomaly interpretation method to attribute detected anomalies. Specifically, after HTS-VAE identifies an anomalous time window, we aim to attribute the anomaly to specific spectral bands and temporal points, thereby pinpointing the sources of change within the dynamic sample. However, a major challenge arises: anomalous observations can contaminate the learned embeddings in HTS-VAE, leading to biased reconstructions even for the normal bands or time points. For example, an anomalous NDVI drop caused by cloud contamination may degrade the reconstruction of temporally adjacent or spectrally related bands such as NIR or Red, making it difficult to precisely localize the source of the change. To attribute detected anomalies to specific spectral bands and time points, we require a method that can isolate individual spectral-temporal dimensions while preserving the context of the remaining data. Full-sample reconstruction may obscure localized signals, whereas dimension-wise analysis enables finer attribution. Gibbs sampling \citep{casella1992explaining} offers an effective solution by iteratively refining one dimension at a time, conditioned on the rest. This allows us to approximate what each anomalous value would be under normal conditions, without disrupting the overall data structure. Based on this, we propose a Gibbs sampling-based anomaly attribution framework to refine the interpretation of detected anomalies by approximating what the “normal” reconstruction should be. The key idea is to iteratively update one spectral-temporal dimension at a time while conditioning on the fixed normal subset, allowing efficient and precise identification of the contribution of each dimension to the anomaly. 

For an input time series of one sample $\textbf{x} \in \mathbb{R}^{C \times T}$, we aim to interpret the detected anomalies by identifying which spectral bands and time points contribute most. After anomaly detection using HTS-VAE, we divide $\textbf{x}$ into a supposedly normal subset $\textbf{x}_n$ and a potentially anomalous subset $\textbf{x}_a$ based on the initial reconstruction score $\text{S}^0$, computed as the negative log-likelihood of $\textbf{x}$ under the learned HTS-VAE distribution. To quantify what is “normal,” we compute a baseline score $b = \frac{1}{N} \sum_{n=1}^{N} \sum_{c,t} \text{S}^0_{c,t}$ as the average reconstruction score over $N$ sliding windows from the training set. $\text{S}^0_{c,t}$ denotes the reconstruction score of band $c$ at time $t$. 

We then perform Gibbs sampling to refine $\textbf{x}_a$. In each Gibbs iteration, instead of replacing multiple anomalous points simultaneously, we sequentially select one spectral band and time point $(c, t)$ identified as anomalous, condition on the fixed normal subset $\textbf{x}_n$ and the rest of $\textbf{x}_a$, and impute the selected dimension to approximate the normal pattern. After updating, we reconstruct the full sample $\textbf{x}'$ and compute its reconstruction score $\text{S}^r$. The iteration proceeds until the updated reconstruction score exceeds the baseline $b$, indicating that the model has recovered a latent embedding corresponding to normal conditions. Finally, we compare the reconstruction score of $\textbf{x}_a$ and $\textbf{x}_a'$ to derive a dimension-wise anomaly attribution score $\mathbf{AS} \in \mathbb{R}^{C \times T}$. Dimensions with higher score improvements (i.e., greater discrepancy before and after imputation) contribute more to the anomaly.

The full procedure is summarized in Algorithm~\ref{alg:anomaly_attribution_gibbs}.

\begin{algorithm}[t]
\caption{Anomaly Attribution using Gibbs Sampling}
\label{alg:anomaly_attribution_gibbs}
\begin{algorithmic}[1]
\REQUIRE Input sample $\textbf{x} \in \mathbb{R}^{C \times T}$, initial reconstruction score $\mathbf{S}^0$, baseline score $b$, maximum iterations $M$
\ENSURE Anomaly attribution score $\mathbf{AS} \in \mathbb{R}^{C \times T}$

\STATE Split $\mathbf{x}$ into supposedly normal $\mathbf{x}_n$ and anomalous $\mathbf{x}_a$ based on $\mathbf{S}^0$
\STATE Initialize $\mathbf{x}' \leftarrow \mathbf{x}$
\FOR{$i = 1$ to $M$}
    \FOR{each anomalous dimension $(c, t)$}
        \STATE Fix $\mathbf{x}_n$ and other dimensions of $\mathbf{x}_a$
        \STATE Sample latent embeddings $(\mathbf{e}_t, \mathbf{e}_s) \sim q_\phi(\mathbf{e}_t, \mathbf{e}_s \mid \mathbf{x}_n \cup \mathbf{x}_a \setminus (c,t))$
        \STATE Reconstruct and impute $\mathbf{x}_a'^{(c,t)}$ conditioned on $(\mathbf{e}_t, \mathbf{e}_s)$
        \STATE Update the $(c,t)$ entry in $\mathbf{x}_a$ with $\mathbf{x}_a'^{(c,t)}$
    \ENDFOR
    \STATE Form full reconstruction $\mathbf{x}' = (\mathbf{x}_n, \mathbf{x}_a)$
    \STATE Compute reconstruction score $\mathbf{S}^r$ from $\mathbf{x}'$
    \IF{$\text{mean}(\mathbf{S}^r) \geq b$}
        \STATE \textbf{break}
    \ENDIF
\ENDFOR
\STATE Compute attribution score $\mathbf{AS} = \mathbf{S}^0 - \mathbf{S}^r$
\RETURN $\mathbf{AS}$
\end{algorithmic}
\end{algorithm}

\subsection{Dynamic sample relabeling}
As shown in Fig. \ref{fig:net} (c), after we get the anomaly patterns and the interpretation results, we can relabel the temporal status of the samples to construct dynamic training samples. Here we consider an attribution significant if the reconstruction likelihood improves by more than a fixed threshold or falls within the top 10\% of all attribution scores. Only significant anomalies are used to guide dynamic sample relabeling, ensuring that relabeling decisions are based on substantial spectral-temporal deviations. We train a lightweight classifier to predict the class label of each time step within a sample, based on the spectral-temporal embeddings derived from the HTS-VAE. This enables us to transform a single static label (e.g., a known wetland class at time $t_0$) into a full sequence of dynamic labels $\{\hat{y}_t\}_{t=1}^{T}$ for the entire time series.

Specifically, for each sample $\mathbf{x} \in \mathbb{R}^{C \times T}$, the HTS-VAE provides its compressed spectral and temporal representations $\mathbf{e}_s$ and $\mathbf{e}_t$. These embeddings, enriched by temporal context and denoised by reconstruction, are concatenated or fused as the feature vector for each time point $t$: $\mathbf{f}_t = \text{Fuse}(\mathbf{e}_{s_t}, \mathbf{e}_{t_t}), \ \text{for } t = 1, \dots, T$. These features $\{\mathbf{f}_t\}$ serve as inputs to a classifier $\mathcal{C}$, which is trained to output the predicted class label $\hat{y}_t$: $\hat{y}_t = \mathcal{C}(\mathbf{f}_t).$ To supervise this classifier, we use the stable label $y_0$ from the known timestamp $t_0$ as the reference class. During training, we assume that all time steps that are not identified as anomalous inherit the same label $y_0$, and only the anomalous time steps require prediction. This semi-supervised setup avoids overfitting and propagates meaningful class transitions only when justified by the anomaly signals. For time steps identified as “normal,” the sample keeps its original label $y_t$. For time steps flagged as anomalous and attributed to meaningful changes, the classifier $\mathcal{C}$ is used to infer their new label $\hat{y}_t$.

Based on this, we successfully achieve "Static to Dynamic": Generating dynamic samples only using existing static annotations. The re-labeled sequences construct a dynamic sample pool, supporting temporal modeling and dynamic land monitoring (i.e., dynamic land cover mapping and wetland mapping).

\section{Study area and material}
\subsection{Study area}
\begin{figure}[h]
    \centering
    \includegraphics[width=0.9\linewidth]{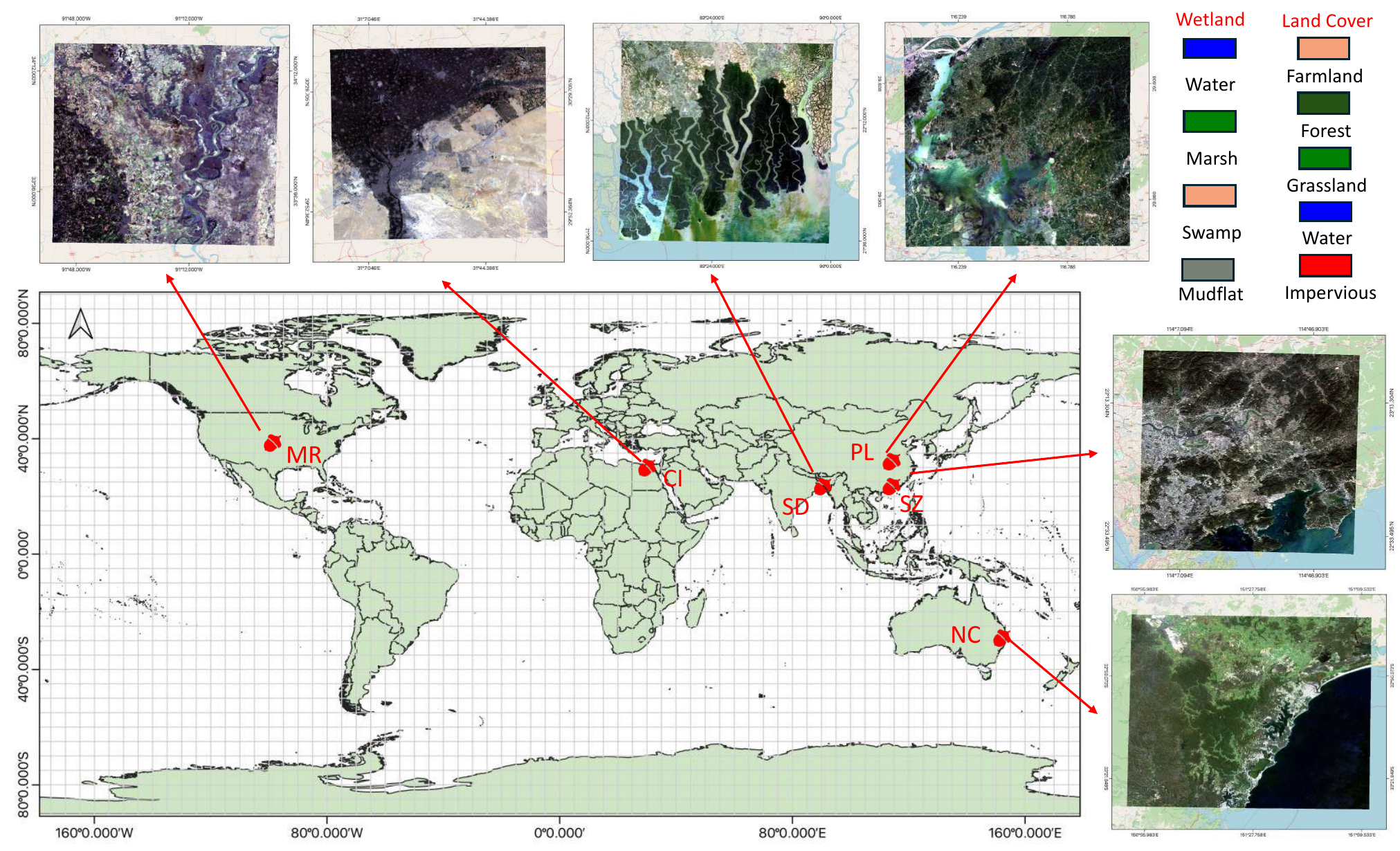}
    \vspace{-1em}
    \caption{The study areas and the classification systems in our study.}
    \label{fig:study area}
\end{figure}

To comprehensively evaluate the effectiveness of the proposed TasGen framework, we conduct experiments on two representative tasks—dynamic land cover mapping and dynamic wetland mapping—across six carefully selected study areas distributed on different continents (Fig. \ref{fig:study area}). Specifically, the dynamic land cover mapping task is conducted in: 1) Shenzhen, China (SZ); 2) Cairo, Egypt (CI); 3) Newcastle, Australia (NC). The dynamic wetland mapping task is evaluated in: 1) Poyang Lake, China (PL); 2) Mississippi River, United States (MR); 3) Sundarbans, India \& Bangladesh (SD).

These regions are chosen based on the following considerations: First, most study areas exhibit a mixture of diverse land cover or wetland types, and the landscapes span different ecological and climatic zones across continents, providing a broad spectrum of surface dynamics. Besides, the areas of these regions are large enough ($12,062 \ km^2 $ of each) to capture significant spatial heterogeneity and representativeness, enabling robust evaluation of dynamic mapping methods at multiple scales. Second, there are distinct temporal evolution patterns across these regions. For example, SZ and CI exhibit rapid urban expansion within short periods, while NC remains relatively stable in terms of land cover. In the wetland mapping task, PL and MR show pronounced seasonal variations in wetland extent and type, whereas SD maintains a relatively stable wetland status throughout the year.

In summary, these six areas capture typical scenarios of land surface dynamics across various ecological and urban environments, featuring both homogeneous and heterogeneous landscapes. This diversity in surface composition and surface change magnitude makes them well-suited for evaluating the land cover mapping and wetland mapping capabilities of TasGen, particularly its ability to automatically adapt to fine-grained temporal-spectral changes and generate temporally consistent classification results. These scenarios also provide a robust benchmark to assess the adaptability and accuracy of large-scale dynamic land cover classification and wetland mapping, especially when relying on automatically relabeled training data to track surface type transitions under complex and rapidly evolving environmental conditions.

\subsection{Material}

In this study, we employ the Global 30 m Seamless Data Cube (SDC30) to support both dynamic land cover mapping and dynamic wetland mapping tasks. SDC30 is a 30 m-resolution, Landsat-like, daily land surface reflectance dataset spanning 2000–2022, generated by fusing observations from Landsat 5, 7, 8, 9, and MODIS Terra, featuring spatiotemporal continuity and high accuracy to support global environmental monitoring \citep{chen2023robot, chen2024global}. The spatiotemporal continuity provides efficiency and flexibility for our model to capture both short-term and long-term dynamics. For each of the six study areas, we use four-day synthetic SDC30 images spanning four years (i.e., 368 observations per area). All synthetic observations are cloud-free, analysis-ready data, with six bands of 30 m resolution (i.e., Blue, Green, Red, NIR, SWIR1, SWIR2). 

Collected SDC30 data and corresponding high-resolution Google Earth images are further used for annotating and validating training samples. Note that here we independently label land cover or wetland types at each timestamp to verify the model's performance. For the land cover mapping task, five major land cover categories were used for classification and change detection: farmland, forest, grassland, water, and impervious surfaces. For the wetland mapping task, the focus is placed on seasonal and long-term variation in four major classes, including surface water, swamp, marsh, and mudflat.

\section{Experiments and evaluation}
\subsection{Experiment settings}
We conduct our experiment on the PyTorch deep learning framework, with 1 NVIDIA GeForce RTX 4090 GPU and 50 training epochs. The batch size is set to 32. The learning rate starts at 0.002 and decreases by a factor of 0.1 after every 10 epochs. We use mini-batch stochastic gradient descent (SGD) as the optimizer for classifier training, and set a momentum of 0.9 and a weight decay of 0.0005. The sliding window length is set to 30. The number of samples is set to 40 for Monte Carlo integration, and the imputation time is set to 10 for Gibbs iteration. 

\subsection{Metrics}
The procedure of the TasGen model comprises two parts: First, the TasGen model detects the anomaly in the time-series data of the sample; Second, based on the detected anomaly, the TasGen model relabels the temporal status of the samples to construct dynamic training samples. After that, dynamic mapping is conducted based on the dynamic training samples. Therefore, the evaluation metrics have three aspects. For anomaly detection evaluation, we mainly use the F1-score (F1-A) to assess the performance. For dynamic sample relabeling, we mainly use the F1-score (F1-S) to assess the performance. For dynamic mapping, we use the overall accuracy (OA) and Kappa coefficients (Kappa) to evaluate the accuracy.

\subsection{Results of TasGen}
TasGen generates dynamic samples via two stages: anomaly detection and dynamic sample relabeling. After getting dynamic samples, dynamic mapping is conducted through a trained classifier. Therefore, we evaluate the performance of TasGen based on anomaly detection, dynamic sample relabeling, and dynamic mapping. We also visualize the performance of the anomaly attribution module in this section.

\begin{figure}[h]
    \centering
    \includegraphics[width=0.8\linewidth]{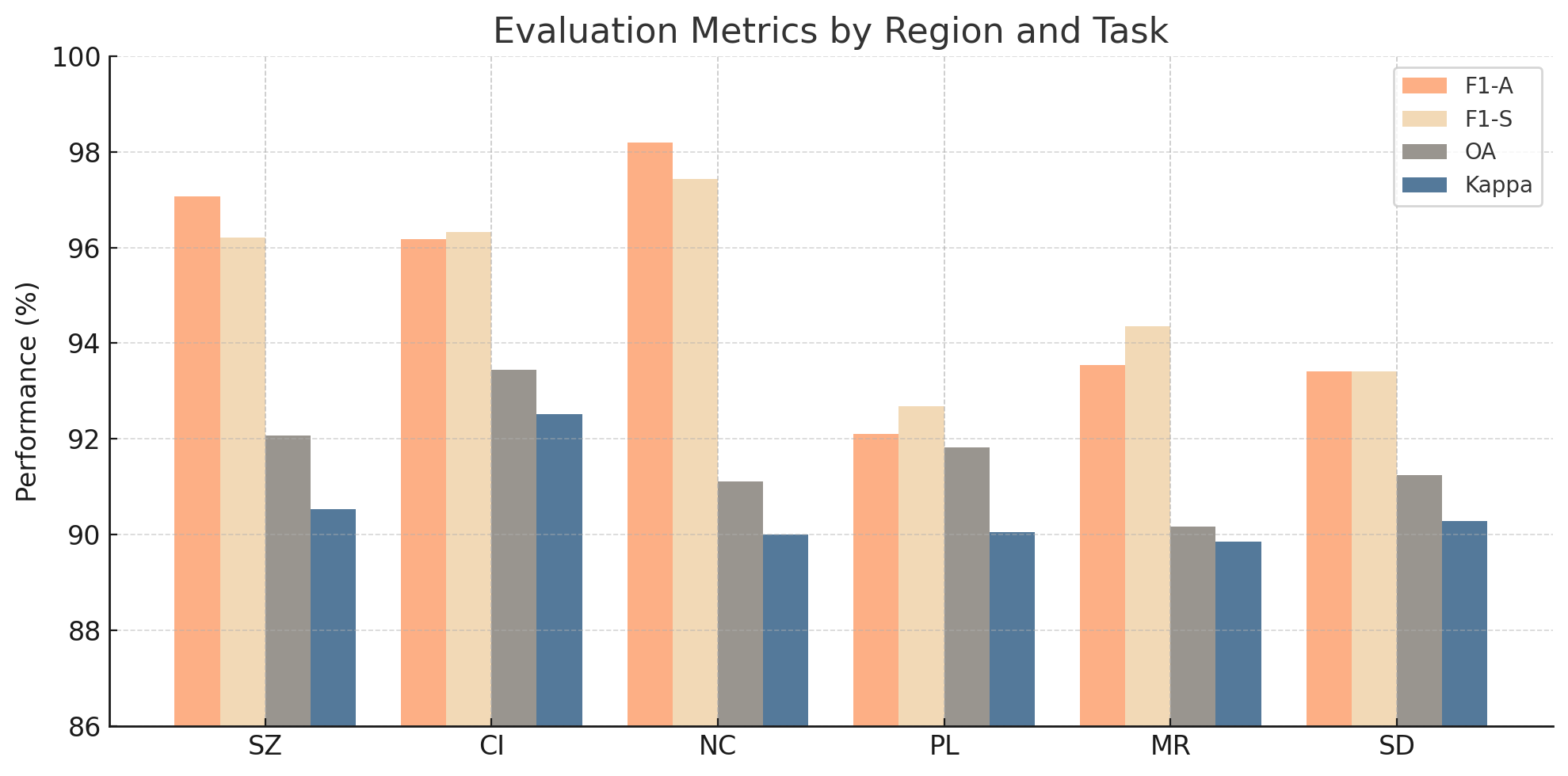}
    \vspace{-1.5em}
    \caption{The results of two tasks in six regions (in \%), including anomaly detection, dynamic sample relabeling and dynamic mapping.}
    \label{fig:evaluation}
\end{figure}

\begin{figure}[t]
    \centering
    \includegraphics[width=0.9\linewidth]{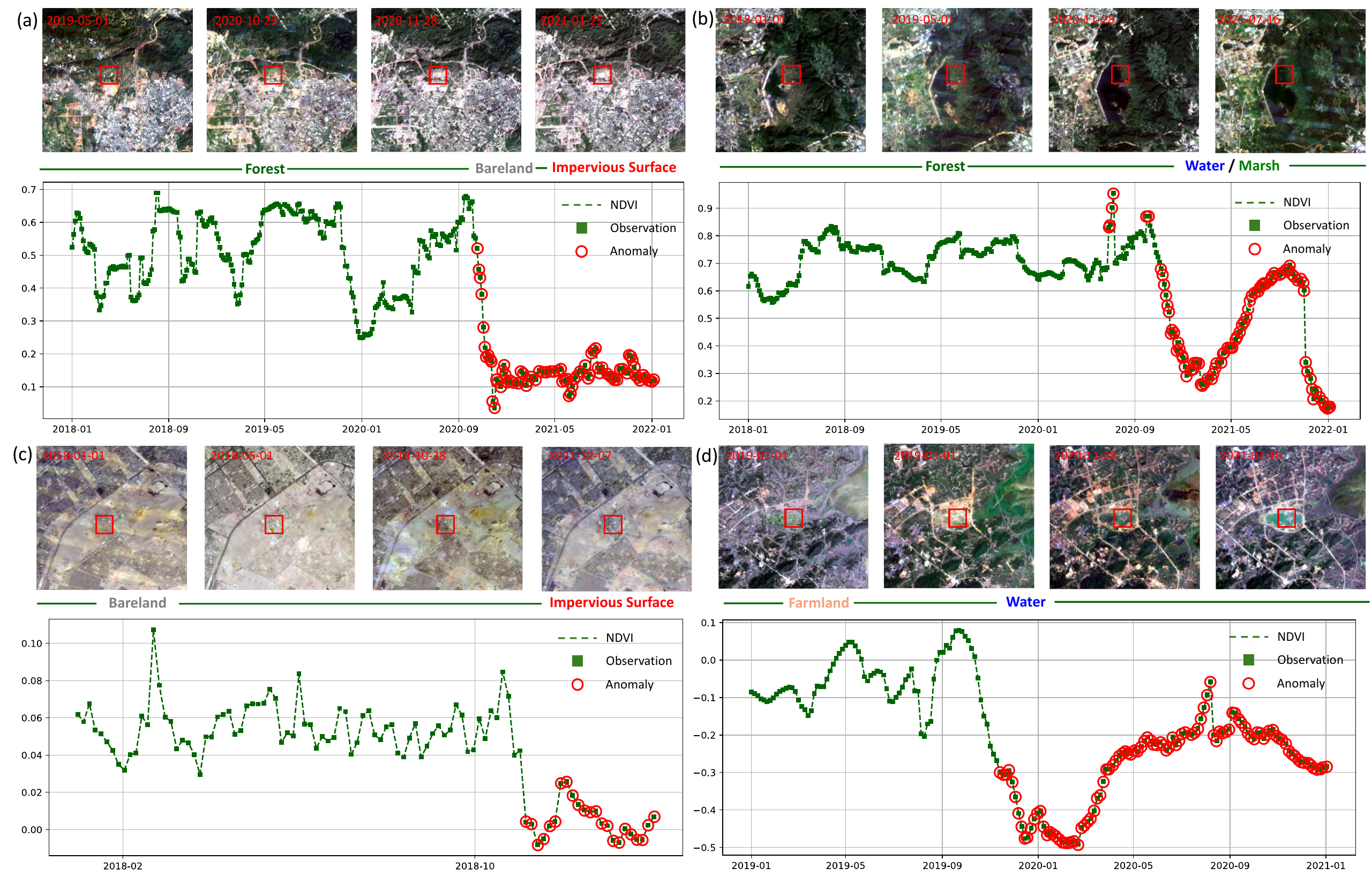}
    \vspace{-1.5em}
    \caption{The visualization results of anomaly detection.}
    \label{fig:anomaly_detection}
\end{figure}

\textbf{Anomaly Detection.} TasGen exhibits excellent performance in detecting temporal anomalies across all regions and mapping tasks. In Fig. \ref{fig:evaluation}, the anomaly detection F1-score (F1-A) consistently exceeds 92\% in all six regions, with the highest reaching 98.19\% in the NC region. This result demonstrates TasGen's strong ability to capture complex spatiotemporal perturbations that signify meaningful land surface changes. Vertically, TasGen achieves higher F1-A scores in the land cover mapping regions (SZ, CI, NC) compared to the wetland mapping regions (PL, MR, SD). This difference may be attributed to the higher temporal variability and spectral ambiguity of wetland environments, where seasonal hydrological dynamics and vegetation transitions introduce subtle yet important variations. Nevertheless, the fact that TasGen still maintains F1-A scores above 92\% in wetland areas proves its robustness across different ecological domains. As Fig. \ref{fig:anomaly_detection} shows, TasGen's hierarchical detection strategy based on spectral-temporal modeling and anomaly attribution offers higher interpretability and finer granularity, resulting in superior anomaly identification.

\textbf{Dynamic sample relabeling.} For dynamic relabeling, TasGen also demonstrates stable and accurate performance. The F1-score for sample relabeling (F1-S) is above 96\% in all land cover regions and above 92\% in all wetland regions. These results confirm the effectiveness of TasGen's two-stage approach: first detecting meaningful sample-level anomalies, then inferring the correct label transitions based on learned temporal-spectral priors. The slight drop in F1-S within wetland regions (e.g., PL: 92.69\%) reflects the intrinsic difficulty of characterizing seasonal wetland state transitions. However, the strong alignment between F1-A and F1-S scores in each region highlights the internal consistency of the model—accurate anomaly detection leads to trustworthy sample relabeling. This consistency is crucial for generating high-quality training samples to support subsequent mapping. In contrast to static sample approaches or heuristic rule-based relabeling (e.g., change-before-label strategies), TasGen's variational inference framework enables learning from data in a probabilistic manner, with built-in uncertainty modeling and sample-level adaptability. This facilitates the generation of flexible, region-adaptive dynamic training sets without requiring manual interventions.

\begin{figure}[h]
    \centering
    \includegraphics[width=0.6\linewidth]{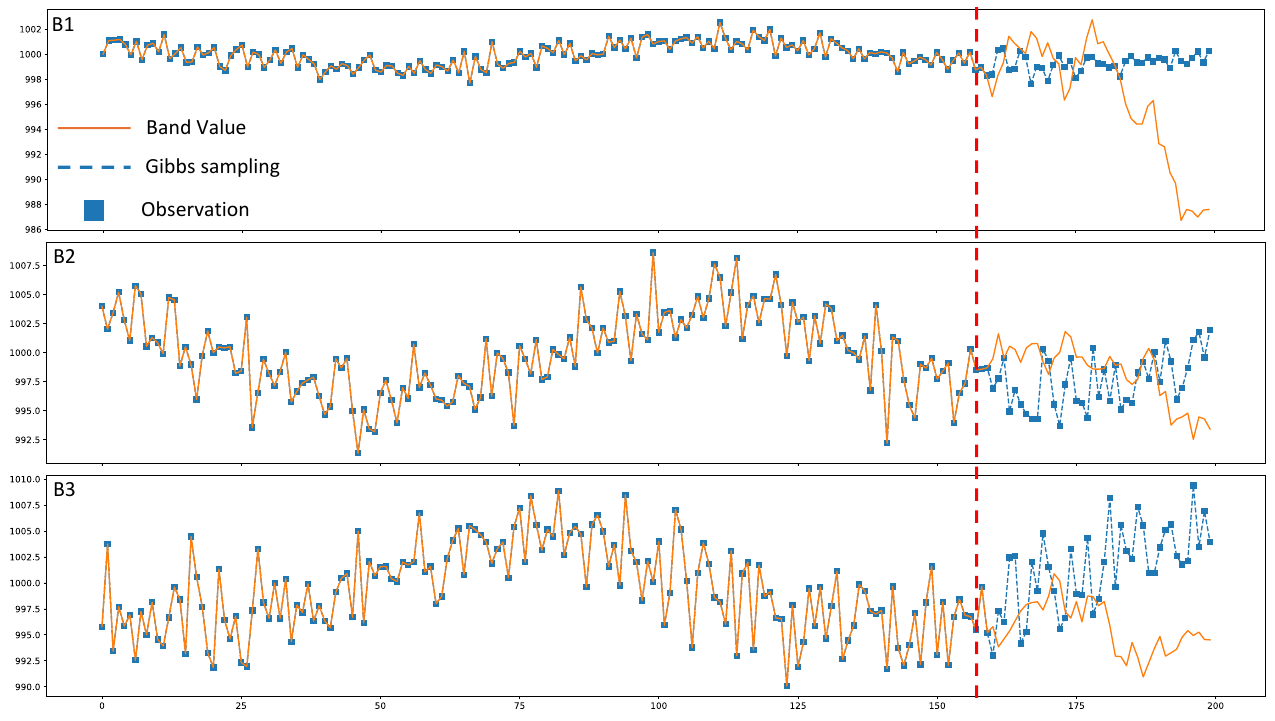}
    \vspace{-1.5em}
    \caption{The results of Gibbs sampling-based anomaly attribution. The attribution score of B1 band is the largest.}
    \label{fig:gibbs}
\end{figure}

\textbf{Anomaly attribution.} As Fig. \ref{fig:gibbs} shows, the Gibbs sampling-based anomaly attribution highlights the most influential dimensions responsible for the detected change. In this example, the attribution score peaks in band B1, indicating that changes in this spectral band are the primary drivers of the detected anomaly. This reflects TasGen's ability to isolate not only when a change occurs, but also which specific spectral channels contribute most significantly to the shift. This improves the model's transparency and provides actionable insights into the nature of land surface transitions.

\begin{figure}[h]
    \centering
    \includegraphics[width=0.9\linewidth]{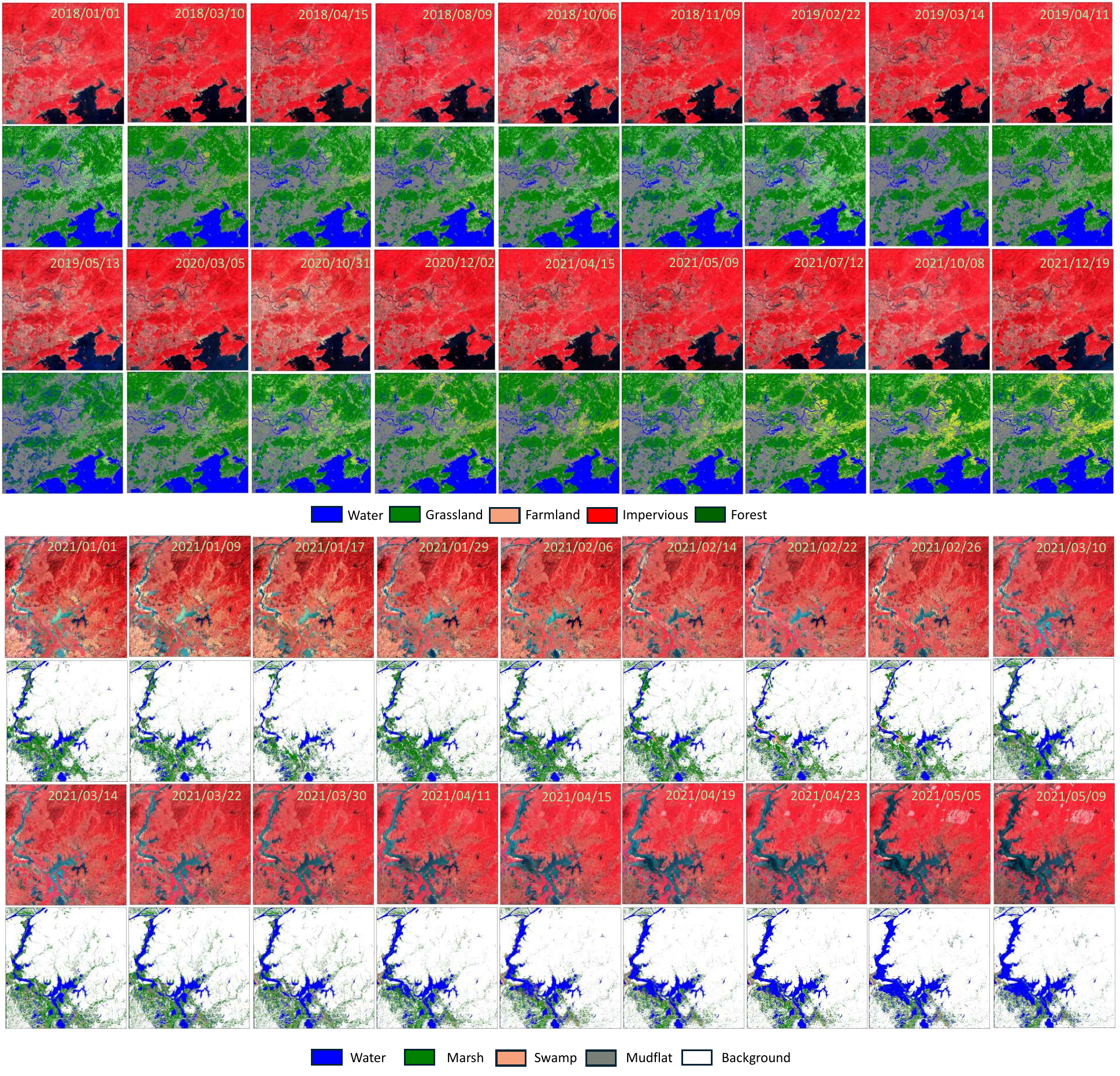}
    \vspace{-1.5em}
    \caption{The visualization results of dynamic mapping performance of TasGen in SZ and PL.}
    \label{fig:result_all}
\end{figure}

\textbf{Dynamic mapping.} For the final dynamic mapping, TasGen achieves high overall accuracy (OA) across all regions, with scores above 89\% in all cases. The Kappa coefficients also remain strong, validating not only the correctness of the predictions but also their statistical consistency with reference labels. Among the land cover regions, the CI region achieves the highest OA (93.44\%) and Kappa (0.925), likely due to clearer land use boundaries and lower intra-class variability. For wetland mapping, the PL region performs best OA (91.82\%), showcasing TasGen’s ability to generalize to hydrologically dynamic and ecologically sensitive zones. From a vertical perspective, the mapping accuracy positively correlates with both F1-A and F1-S scores, reinforcing the importance of high-quality sample generation. TasGen’s ability to detect, correct, and relabel sample-level dynamics translates effectively into pixel-level classification outcomes. Fig. \ref{fig:result_all} shows the visualization results of dynamic mapping.

\subsection{Comparative studies}

To evaluate the effectiveness of TasGen in comparison with existing remote sensing time-series analysis methods, we conduct a comprehensive comparison against four widely-used baselines: CCDC \citep{zhu2014continuous}, LandTrendr \citep{kennedy2010detecting}, TSSCD \citep{he2024time}, and static samples. CCDC models pixel-wise temporal trends using harmonic regression to continuously detect land surface changes and classify land cover types at each change point. LandTrendr segments spectral time series into linear pieces to detect long-term trends of disturbance and recovery. TSSCD leverages temporal dense annotations to model the temporal patterns using efficient deep learning networks, to perform semantic segmentation of land cover changes using long-range temporal context. Static samples use manually labeled static samples to train a classifier, assuming label consistency over time and lacking adaptability to dynamic land surface changes. Each method is applied to the same six regions and evaluated on the final dynamic mapping performance using Overall Accuracy (OA) and Kappa coefficient. Table \ref{tab:oa} and Table \ref{tab:kappa} show the comparison results.

\begin{table}
\centering
\caption{Comparison of mapping accuracy (OA in \%) across six regions.}
\label{tab:oa}
\resizebox{\linewidth}{!}{
\begin{tabular}{ccccccc}
\hline
\textbf{Task} & \textbf{Region} & \textbf{LandTrendr} & \textbf{CCDC} & \textbf{TSSCD} & \textbf{Static} & \textbf{TasGen (Ours)} \\
\hline
\multirow{3}{*}{\textbf{Dynamic Land Cover Mapping}} 

&    SZ & 85.91 ± 0.36 & 87.64 ± 0.37 & 91.93 ± 0.31 & 85.71 ± 0.46 & \textbf{92.07 ± 0.39} \\
&    CI & 86.01 ± 0.35 & 89.49 ± 0.51 & 92.83 ± 0.53 & 85.57 ± 0.37 & \textbf{93.44 ± 0.43} \\
&    NC & 88.74 ± 0.29 & 87.58 ± 0.23 & \textbf{92.17 ± 0.39} & 86.31 ± 0.36 & 91.11 ± 0.61 \\
    \hline
\multirow{3}{*}{\textbf{Dynamic Wetland Mapping}}
&    PL & 83.35 ± 0.62 & 84.12 ± 0.69 & 88.14 ± 0.61 & 82.19 ± 0.54 &  \textbf{91.82 ± 0.70} \\
&    MR & 83.86 ± 0.45 & 85.46 ± 0.56 & 89.64 ± 0.64 & 81.43 ± 0.35 & \textbf{90.17 ± 0.79} \\
&    SD & 84.48 ± 0.45 & 85.64 ± 0.67 & 90.01 ± 0.58 & 80.82 ± 0.56 & \textbf{91.25 ± 0.54} \\
\hline
\end{tabular}}
\end{table}

\begin{table}
\centering
\caption{Comparison of mapping accuracy (Kappa) across six regions.}
\label{tab:kappa}
\resizebox{\linewidth}{!}{
\begin{tabular}{ccccccc}
\hline
\textbf{Task} & \textbf{Region} & \textbf{LandTrendr} & \textbf{CCDC} & \textbf{TSSCD} & \textbf{Static} & \textbf{TasGen (Ours)} \\
\hline
\multirow{3}{*}{\textbf{Dynamic Land Cover Mapping}} 

&    SZ & 85.01 ± 0.36 &  86.35 ± 0.21 &  90.14 ± 0.50 & 84.05 ± 0.36 & \textbf{90.54 ± 0.59} \\
&    CI &  85.61 ± 0.45 & 88.02 ± 0.37 & 92.01 ± 0.63 & 85.47 ± 0.27 & \textbf{92.52 ± 0.53} \\
&    NC & 87.82 ± 0.39 & 87.32 ± 0.43 & \textbf{91.46 ± 0.58} & 86.02 ± 0.36 & 91.01 ± 0.31 \\
\hline
\multirow{3}{*}{\textbf{Dynamic Wetland Mapping}}
&    PL & 83.13 ± 0.32 & 83.69 ± 0.69 &  87.85 ± 0.65 & 82.08 ± 0.54 &  \textbf{90.05 ± 0.63} \\
&    MR & 83.37 ± 0.41 & 85.22 ± 0.36 & 89.23 ± 0.53 & 80.79 ± 0.45 & \textbf{89.86 ± 0.60} \\
&    SD &  84.12 ± 0.25 & 85.21 ± 0.47 & 89.61 ± 0.54 & 80.16 ± 0.46 & \textbf{90.29 ± 0.64} \\
\hline
\end{tabular}}
\end{table}

\begin{figure}[t]
    \centering
    \includegraphics[width=0.95\linewidth]{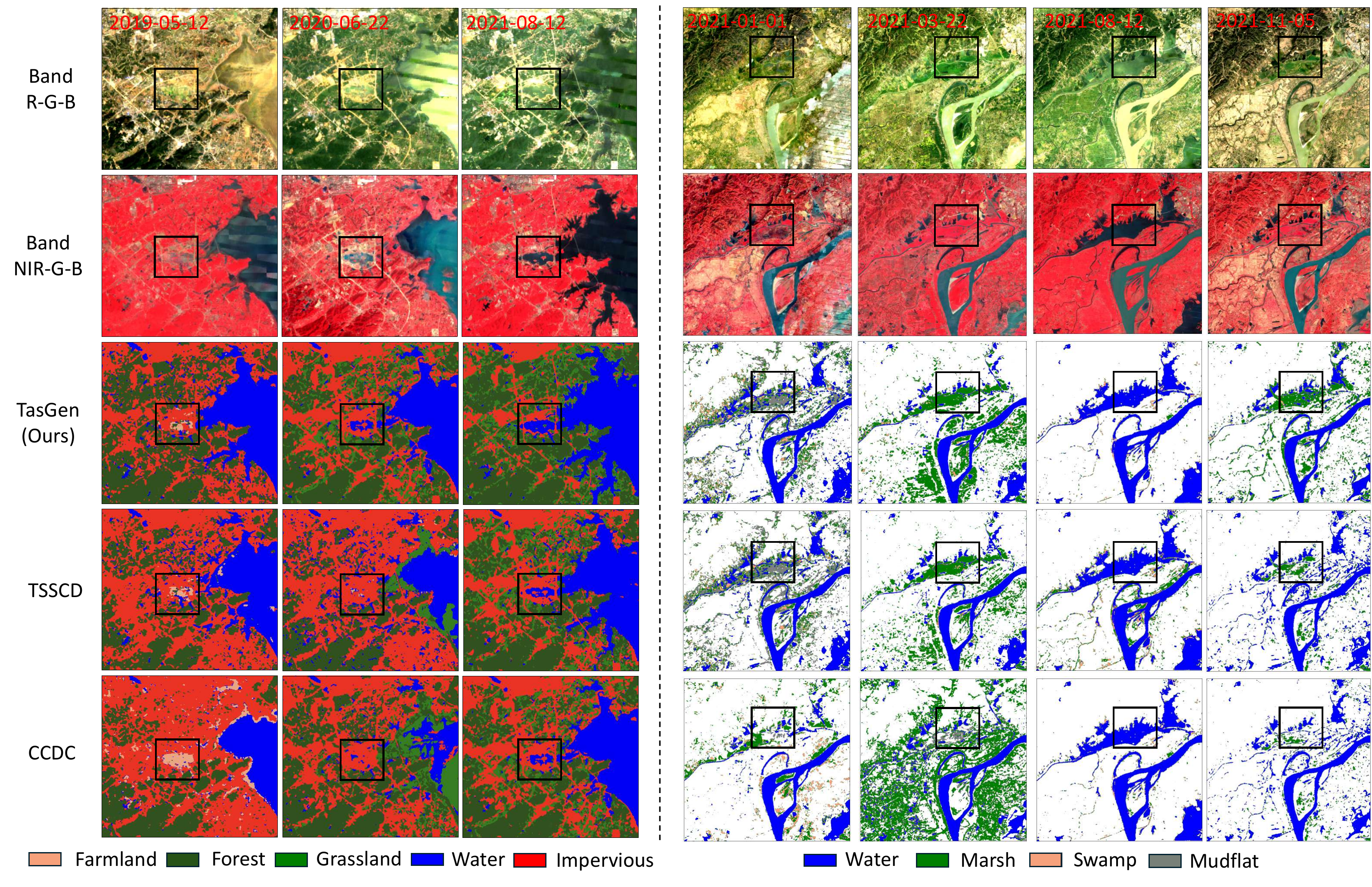}
    \vspace{-1.5em}
    \caption{The visualization results of dynamic mapping, with the comparison with TSSCD and CCDC.}
    \label{fig:comparison_1}
\end{figure}
\begin{figure}[h]
    \centering
    \includegraphics[width=0.8\linewidth]{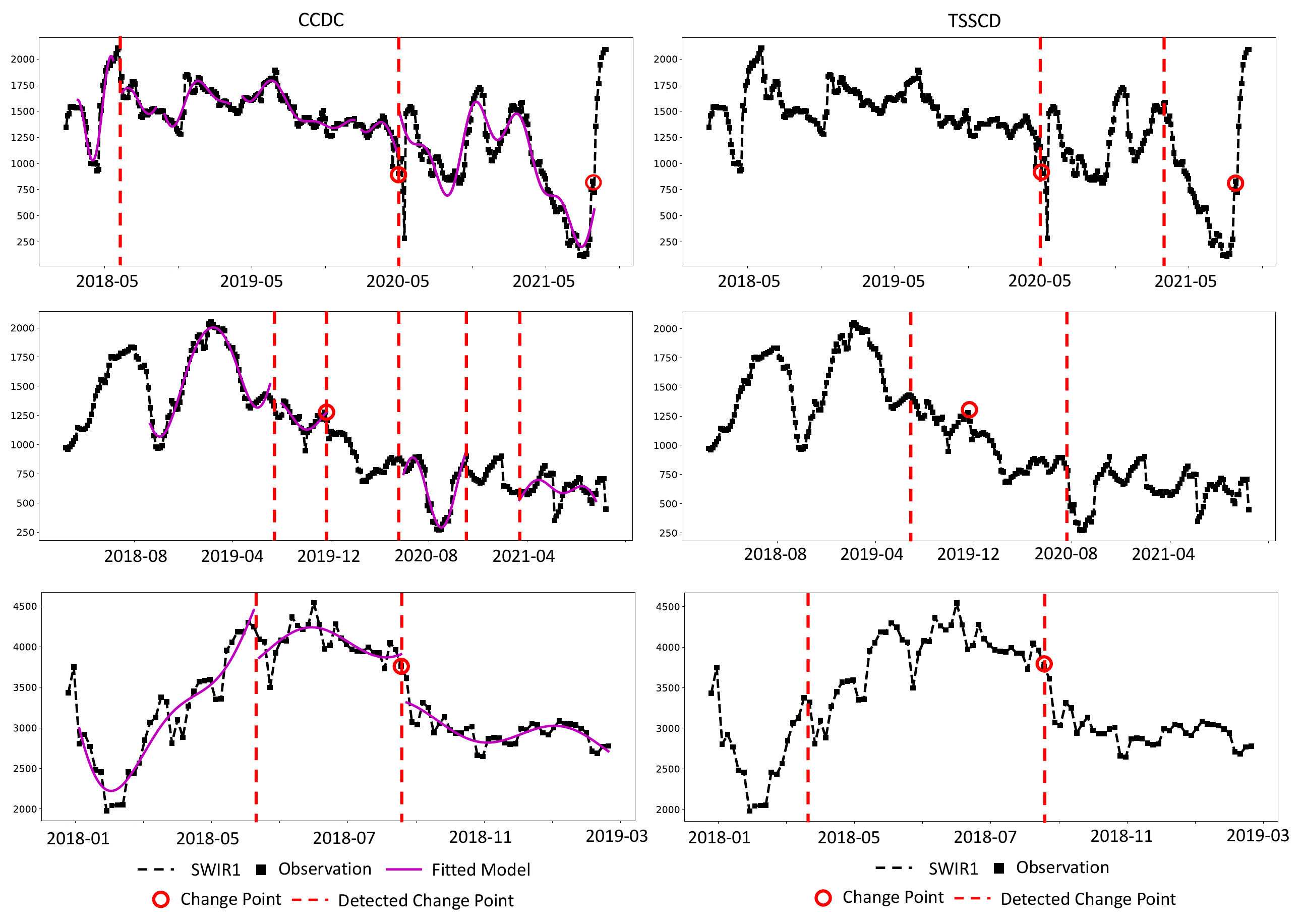}
    \vspace{-1.5em}
    \caption{The visualization results of change detection performance of TSSCD and CCDC, with the same areas of (b)-(d) as the Fig. \ref{fig:evaluation}}
    \label{fig:comparison_2}
\end{figure}

TasGen achieves the best OA and Kappa in almost regions, with particularly large margins in complex wetland areas (PL, MR, SD), where static or change detection-based methods struggle to adapt to seasonal hydrological dynamics. This highlights the value of dynamic sample generation and anomaly-aware relabeling. Compared to CCDC and LandTrendr, which model pixel-level changes but ignore sample-level class transitions, TasGen leverages learned spectral-temporal embeddings and anomaly reasoning to reassign class labels precisely at the sample level. Although TSSCD achieves strong results, it relies on dense manual annotations and lacks sample-level interpretability. TasGen provides a more flexible, explainable, and annotation-efficient alternative, bridging change detection and supervised mapping. Static sample-based methods consistently underperform, reinforcing the need for dynamic modeling in rapidly evolving surface environments. Fig. \ref{fig:comparison_1} visualizes the dynamic mapping results of TasGen, TSSCD and CCDC. Fig. \ref{fig:comparison_2} compares the change detection performance of TSSCD and CCDC in the same areas as Fig. \ref{fig:anomaly_detection} (b)–(d). Both methods exhibit notable misdetections and false alarms, especially under complex surface dynamics such as wetland transitions, compounded by dense observation frequencies (e.g., 4-day intervals). In contrast, although TasGen may also generate detection errors in the first stage, its subsequent sample relabeling module effectively mitigates these issues by the trained classifier. Therefore, TasGen can prioritize high recall during anomaly detection to ensure comprehensive change capture, while leveraging the second-stage relabeling to refine accuracy and reduce false positives. This comparison confirms that TasGen offers superior performance by trustable anomaly detection and accurate sample relabeling. Introducing dynamic samples as an interpretable and controllable intermediate variable increases the fault tolerance of anomaly detection and improves the accuracy of dynamic mapping.

\section{Discussion}
\subsection{Effectiveness of modules in anomaly detection}
To evaluate the individual contribution of each module within TasGen, we conduct an ablation study across six geographically and ecologically diverse regions. Table \ref{tab:ablation-transposed} summarizes the anomaly detection F1-score (F1-A) under five model settings: the full TasGen architecture and four ablated variants where one key module is removed at a time.

\textbf{Rationale for decoupling and coupling.} Removing the \textit{dual-dimension embedding} means we directly learn the temporal and spectral features jointly, without decoupling them in the first stage. Removing the \textit{HTS-VAE} means we learn the temporal and spectral patterns separately, without recognizing them as a whole in the reconstruction process. We can see from Table \ref{tab:ablation-transposed} that removing the \textit{dual-dimension embedding} structure leads to a significant performance drop. This module enables explicit compression of temporal and spectral dependencies into two disentangled embeddings, which are then fused hierarchically. Without this disentanglement, the model suffers from representational entanglement: temporal fluctuations may interfere with spectral pattern recognition, and vice versa. This makes it harder for the model to isolate true anomalies from natural variation in either domain. Disabling the \textit{HTS-VAE} results in the largest drop in most regions. HTS-VAE introduces a structured generative process, where temporal context $\textbf{e}_t$ guides the learning of spectral representations $\textbf{e}_s$, ensuring that spectral anomalies are interpreted within their correct temporal phase. Without hierarchical conditioning, the model treats spectral and temporal embeddings independently, which leads to temporal misalignment and inaccurate anomaly detection.

\begin{table}[h]
\caption{Anomaly detection performance (F1-A, in \%) under different ablation settings across six regions.}
\label{tab:ablation-transposed}
\centering
\resizebox{\linewidth}{!}{
\begin{tabular}{ccccccc}
\hline
Task & Region & w/o Dual-dimension & w/o HTS-VAE & w/o Curation & w/o Glow & Full (Ours) \\
\hline
\multirow{3}{*}{\textbf{Dynamic Land Cover Mapping}} 
& SZ & 94.52 & 93.74 & 95.08 & 96.06 & \textbf{97.06} \\
& CI & 93.67 & 92.98 & 93.87 & 95.23 & \textbf{96.18} \\
& NC & 93.03 & 92.56 & 96.25 & 97.12 & \textbf{98.19} \\
\hline
\multirow{3}{*}{\textbf{Dynamic Wetland Mapping}} 
& PL & 87.26 & 85.39 & 87.13 & 90.07 & \textbf{92.11} \\
& MR & 88.13 & 86.72 & 88.62 & 91.72 & \textbf{93.54} \\
& SD & 87.88 & 86.01 & 89.08 & 91.66 & \textbf{93.42} \\
\hline
\end{tabular}}
\end{table}

\textbf{Rationale for sample curation.} The \textit{sample curation strategy} ensures spectral embeddings are not learned from noisy temporal patterns by reconstructing a clean time-series via Deconv1D layers. Removing this module means the model directly learns the patterns through the raw inputs, introducing risks of overfitting to clouds, sensor noise, or short-term fluctuations. This is especially problematic in wetland regions (e.g., PL, SD), where false positives are easily triggered by seasonal water-level shifts. The F1-A drops in these cases highlight the importance of pre-filtering as a noise-robust foundation for subsequent representation learning. 

\textbf{Rationale for Glow.} Replacing the \textit{Glow flow-based posterior} with a standard Gaussian reduces model flexibility, resulting in moderate F1-A degradation. The Glow module expands the representational capacity beyond a simple Gaussian assumption, allowing the model to better approximate complex and multimodal latent distributions, which are common in real-world spatiotemporal data.

The \textit{full TasGen model} consistently outperforms all ablated variants in every region, demonstrating the effectiveness of its integrated design. This ablation study confirms the necessity of each component in the anomaly detection stage of TasGen. Dual-dimensional embedding and HTS-VAE ensure structurally aligned temporal-spectral modeling, sample curation suppresses spurious signals, and Glow strengthens posterior modeling. Together, they form a synergistic framework for reliable anomaly detection for dynamic sample generation later.

\subsection{Effectiveness of feature design in dynamic sample relabeling} 

To further assess the impact of different feature designs on dynamic sample relabeling, we conduct an ablation study by replacing the default feature $\mathbf{f}_t = \text{Fuse}(\mathbf{e}_{s_t}, \mathbf{e}_{t_t})$ used by the classifier $\mathcal{C}$ with several alternative representations. This allows us to validate whether the proposed hierarchical temporal-spectral embeddings learned by HTS-VAE are indeed the most effective representation for label propagation across time steps.

We compare the following variants: (1) Full HTS-VAE Fusion (Ours): The default setting, where spectral and temporal embeddings are concatenated after denoising and hierarchical encoding; (2) Only Temporal Embedding ($\textbf{e}_t$): Using only the temporal embedding, ignoring inter-band spectral dependencies; (3) Only Spectral Embedding ($\textbf{e}_s$): Using only the spectral embedding, discarding the learned temporal dynamics; (4) Raw Feature ($\textbf{x}_t$): Directly using the raw spectral vector at each time $t$ without any embedding or context modeling.

The results in Table \ref{tab:relabeling-ablation} show the relabeling F1-score (F1-S) across all six regions. The results clearly indicate that the full fusion of temporal and spectral embeddings yields the highest relabeling accuracy in all regions, confirming the necessity of joint modeling. Specifically, using only $\textbf{e}_t$ or $\textbf{e}_s$ causes a noticeable drop in F1-S, showing that neither temporal nor spectral information alone is sufficient to resolve complex class transitions, particularly in areas with both phenological variation and spectral ambiguity (e.g., wetlands like PL and SD). Using raw features without embedding consistently leads to inferior results. This suggests that TasGen’s learned embeddings not only reduce dimensionality but also encode meaningful temporal-spectral context, making them more discriminative for downstream classification. The rule-based interpolation method performs worst, as it cannot generalize beyond simple linear trends or handle non-symmetric transitions (e.g., permanent conversion of land use). 

\begin{table}[h]
\caption{Ablation study on dynamic sample relabeling accuracy (F1-S, in \%) across six regions using different feature designs.}
\label{tab:relabeling-ablation}
\centering\resizebox{0.7\linewidth}{!}{
\begin{tabular}{cccccc}
\hline
Task & Region & Only $\textbf{e}_t$ & Only $\textbf{e}_s$ & Raw $\textbf{x}_t$ & Full Fusion\\
\hline
\multirow{3}{*}{\textbf{Dynamic Land Cover Mapping}} 
& SZ & 92.88 & 93.02 & 90.14 & \textbf{96.21}\\
& CI  & 92.52 & 93.19 & 89.80 & \textbf{96.33}\\
& NC  & 94.24 & 95.31 & 91.76 & \textbf{97.43}\\
\hline
\multirow{3}{*}{\textbf{Dynamic Wetland Mapping}} 
& PL  & 88.56 & 89.42 & 86.03 & \textbf{92.69}\\
& MR  & 90.81 & 91.47 & 87.12 & \textbf{94.35}\\
& SD  & 91.22 & 91.96 & 88.63 & \textbf{93.41}\\
\hline
\end{tabular}}
\end{table}

These results validate that the joint embedding learned by HTS-VAE not only improves anomaly detection, but also forms a powerful foundation for sample-level dynamic label inference. The dual encoding mechanism supports both contextual awareness and noise robustness, essential for scalable time-series mapping in real-world remote sensing applications.

\subsection{Robustness test}

To evaluate the robustness of TasGen under real-world remote sensing challenges such as irregular observation frequency and temporal data loss, we design two complementary experiments: (1) temporal frequency reduction, and (2) random missing observations. These experiments aim to simulate practical scenarios where real satellite observations may be sparse or corrupted due to weather conditions, revisit limitations, or sensor errors.

\textbf{Temporal frequency reduction.} In this setting, we uniformly reduce the number of observations per year from the original 92 (i.e., one observation every four days) to simulate sparser sampling conditions. We consider 1/2 observations (every 8 days), 1/4 observations (every 16 days) and 1/8 observations (every 32 days). As shown in Table \ref{tab:robustness}, TasGen exhibits resilience somehow: even when the temporal frequency is halved, the anomaly detection performance only slightly drops, maintaining F1-A scores above 90\% in most regions. When the threshold achieves 1/8, the anomaly detection performance drops dramatically. This demonstrates the robustness of the temporal embedding design and dual-view representation, allowing the model to learn long-range temporal patterns from limited data to some extent. 

\textbf{Random missing observations.} In this experiment, we randomly mask a percentage of the time points in each sample to simulate common missing scenarios such as cloud occlusion or sensor failure. Specifically, we consider missing ratios of 10\%, 30\%, and 50\%. The masked entries are linearly interpolated, making the final synthesis observations consistent with our training setup. As shown in the Table \ref{tab:robustness}, while the performance of all regions declines with increasing missing rate, TasGen still retains high F1-A scores (over 83\%) with 30\% missing and nearly above 80\% under 50\% missing, highlighting its robustness to incomplete temporal sequences. 

\begin{table}[h]
\caption{Robustness test on anomaly detection F1-score (F1-A, \%) under reduced observation frequency (1/2 obs, 1/4 obs, 1/8 obs) and random missing observations (10\%, 30\%, 50\%).}
\label{tab:robustness}
\centering
\centering
\resizebox{0.92\linewidth}{!}{
\begin{tabular}{cc|cccc|ccc}
\hline
Task &Region & Full obs & 1/2 obs & 1/4 obs & 1/8 obs  & 10\% miss & 30\% miss & 50\% miss\\
\hline
\multirow{3}{*}{\textbf{Dynamic Land Cover Mapping}} 
& SZ  & 97.06 & 95.88 & 94.41 & 84.22 & 95.42 & 90.78 & 87.51 \\
& CI  & 96.18 & 94.33 & 91.74 & 82.15 & 94.17 & 89.11 & 85.72 \\
& NC  & 98.19 & 96.47 & 94.28 & 83.04 & 96.65 & 90.92 & 86.63 \\
\hline
\multirow{3}{*}{\textbf{Dynamic Wetland Mapping}} 
& PL  & 92.11 & 89.80 & 86.23 & 77.76 & 89.03 & 83.14 & 79.44 \\
& MR  & 93.54 & 91.22 & 87.91 & 78.35 & 90.68 & 82.39 & 80.60 \\
& SD  & 93.42 & 91.08 & 88.14 & 78.67 & 90.91 & 83.84 & 81.27 \\
\hline
\end{tabular}}
\end{table}

\section{Conclusion}
To enable scalable, interpretable, and accurate dynamic geographic mapping with minimal manual effort, in this study, we propose TasGen, a novel two-stage framework for automatic dynamic sample generation in time-series dynamic remote sensing mapping. By first decoupling and then jointly modeling spectral and temporal dependencies through a dual-dimension embedding and a hierarchical temporal-spectral variational autoencoder (HTS-VAE), with performing interpretable anomaly attribution using Gibbs sampling, and relabeling the dynamic samples via a trained classifier, TasGen achieves "Static to Dynamic" with high interpretability and reliability: effectively transforming static annotations into temporally dynamic training samples for dynamic mapping with no more human intervention. Extensive experiments across six geographically diverse regions and two mapping tasks, i.e., land cover and wetland mapping, demonstrate that TasGen consistently achieves an average OA of 95.08\%, 95.01\%, and 91.64\% in anomaly detection, dynamic sample generation, and dynamic mapping, respectively, outperforming existing state-of-the-art methods. Based on stable classification theory \citep{gong2024stable}, generated dynamic samples can serve as reliable supervision for high-quality dynamic mapping, which is also proven in our experiments. Comprehensive ablation studies illustrate the effectiveness of modules and designs. The method shows strong robustness to observation sparsity and missing data, while maintaining high adaptability in different surface dynamics. By reducing the dependence on continuous manual annotations and enhancing the temporal validity of training data, TasGen provides a scalable and accurate solution for dynamic remote sensing monitoring, offering new possibilities for large-scale land change analysis and environmental applications.

\bibliography{mybibfile}

\end{document}